\documentclass[10pt,journal,letterpaper,compsoc]{IEEEtran}

\usepackage[dvips]{graphicx}
\DeclareGraphicsExtensions{.eps,.ps}

\usepackage[tight,footnotesize]{subfigure}

\usepackage{algorithm}
\usepackage{algorithmic}
\usepackage{url}

\renewcommand{\thesection}{\Roman{section}}
\newtheorem{theoremm}{Theorem}
\newtheorem{eqed}{Example}
\newtheorem {lemmaa}{Lemma}

\newtheorem {defnn}{Definition}
\newtheorem {corollaryy}{Corollary}
\newtheorem {conjecturee}[theoremm]{Conjecture}

\newtheorem {procd}{Procedure}

\newenvironment{example}{\begin{eqed} \rm}{\hfill\end{eqed}}

\newenvironment{lemma}{\begin{lemmaa} \sl}{\end{lemmaa}}
\newenvironment{theorem}{\begin{theoremm}{\bf :}\sl}{\end{theoremm}}
\newenvironment{defn}{\begin{defnn} \sl}{\end{defnn}}

\begin{document}

\title{Analysis and synthesis of nonlinear reversible cellular automata in linear time}

\author{
Sukanta~Das ~ and ~ Biplab ~K ~Sikdar
\IEEEcompsocitemizethanks{\IEEEcompsocthanksitem Sukanta Das is with the Department of Information Technology, and
Biplab K Sikdar is with the Department of Computer Science \& Technology of
        Bengal Engineering and Science University, Shibpur,
        Howrah, West Bengal, India 711103. email:
    \{sukanta@it,biplab@cs\}.becs.ac.in}\protect\\
\thanks{This research work is partially supported by AICTE career award fund (F.No. 1-51/RID/CA/29/2009-10)}}


\IEEEcompsoctitleabstractindextext{%
\begin{abstract}
Cellular automata (CA) have been found as an attractive modeling tool for various applications, such as, pattern recognition, image processing, data compression, encryption, and specially for VLSI design \& test.
For such applications, mostly a special class of CA, called as linear/additive CA,
have been utilized. Since linear/additive CA refer a limited number of candidate CA, while searching for solution to a problem, the best result may not be expected. The nonlinear CA can be a better alternative to linear/additive CA for achieving desired solutions in different applications. However, the nonlinear CA are yet to be characterized to fit the design for modeling an application. This work targets characterization of the nonlinear CA to utilize the huge search space of nonlinear CA while developing applications in VLSI domain. An analytical framework is developed to explore the properties of CA rules. The characterization is directed to deal with the reversibility, as the reversible CA are primarily targeted for VLSI applications. The reported characterization enables us to design two algorithms of linear time complexities -- one for identification and  nother for synthesis of nonlinear reversible CA. Finally, the CA rules are classified into 6 classes for developing further efficient synthesis algorithm.
\end{abstract}

}
\maketitle

\section{Introduction}
\IEEEPARstart{S}ince the invention of homogeneous
structure of cellular automata (CA) \cite{Neuma66}, it has been employed for modeling
physical systems. To get better insight into a physical system, the CA structure is simplified with a restriction to local interactions among the cells
\cite{Wolfr83}.
The simplified structure, proposed in \cite{Wolfr83}, is an 1-dimensional CA, each cell having two states (0/1)
with uniform 3-neighborhood (self, left neighbor and right neighbor) dependencies among
the cells. It effectively introduces the modularity in a CA structure.

Though, in a number of works \cite{Wolframbook}, it has been shown that the 1-dimensional
3-neighborhood CA exhibit excellent performance while modeling physical systems,
it is hard to view that the interacting objects in a dynamical
system obey the same local rule (homogeneity) during its evolution.
To model such a variety of physical systems, non-homogeneous CA structure
(also called {\em hybrid CA}) is evolved as an alternative choice.
A number of researchers have, therefore, focused their attention to hybrid CA
\cite{Cattel96,ppc1,Horte89a,Nandi94a,Chowd94b} since 1980s and explored the potential design with
1-dimensional hybrid CA specially in $VLSI$ design and test \cite{ppc1,Nandi94a,Chowd94b}.

{T}{he} $PRPG$s (pseudo-random pattern generators), for example, employed for designing the test logic of $VLSI$ circuits, are traditionally implemented with Linear Feedback Shift Registers
($LFSR$s) \cite{Barde87} or with its variations like $GLFSR$ \cite{Glfsr}, Phase-Shift $LFSR$ \cite{Rajski3}, and ring generators \cite{Mruga03,Mrugalski06}. However, the CA based PRPG design have been attracted the researches as an excellent alternative since 1980s \cite{Barde90,Horte89a,Rajski00,Danial97,biplabtcad}. Reversible CA have been a choice for such designs.

The detail characterization of hybrid CA and their applications in $VLSI$ design and test
have been reported in \cite{ppc1}.
All such applications are developed around the linear/additive CA structure.
However, the linear/additive CA refer a limited number of candidate CA while modeling an application.
Therefore, for effective modeling of applications from diverse fields including VLSI design and test, the nonlinear CA can be a better alternative \cite{jetta,tcad/DasS10}. This demands massively available results on nonlinear CA characterization.

Due to unavailability of characterization tool, the nonlinear CA could not be properly characterized. The lack of characterization on nonlinear CA behavior (specially for the reversible CA) links widespread acceptance of linear CA in exploring CA based solutions in VLSI domain. 

The above scenario motivates us to undertake the research for characterization of nonlinear hybrid CA,
targeting $VLSI$ design and test.
An explicit characterization of nonlinear reversible CA, with an attention to $VLSI$ design and test is reported in this paper. The major contributions of the current work can be summarized as:
\begin{itemize}
\item A discrete tool, namely {\em Reachability tree}, has been proposed to characterize nonlinear CA. The tool has been proved very effective to discover 1-dimensional two-state 3-neighborhood CA behavior.
\item An algorithm is proposed to identify, in $O(n)$ time whether a given $n$-cell CA is reversible.
\item For the synthesis of an $n$-cell reversible CA, an $O(n)$ time algorithm is developed.
\item The CA rules, capable of forming reversible CA, are classified into six classes. The relationship among these classes is established. That further simplifies the proposed analysis and synthesis schemes.
\end{itemize}
The preliminary version of this characterization has been reported in \cite{Acri04,Acri06}.
In the subsequent sections, we refer characterization of individual cell rule and the CA as a whole.
To facilitate such characterization of CA, the basics of cellular automata is introduced next.

\section{Cellular Automata Basics}
\label{CAbasics}

A cellular automaton (CA) consists of a number of cells organized in the form of a lattice.
It evolves in discrete space and time, and can be viewed as an autonomous
finite state machine ($FSM$).
Each cell of a CA stores a discrete variable at time $t$ that refers to the present state of the cell.
The next state of the cell at $(t+1)$ is affected by its state and
the states of its $neighbors$ at time $t$.
In this work, we concentrate on one-dimensional 3-neighborhood (self, left and right neighbors) CA,
where a CA cell is having two states - 0 or 1.
In such a CA, the  next state $S^{t+1}_i$ of the $i^{th}$ cell is specified by the
{\em next state function} $f_i$ as
\begin{equation}
S^{t+1}_i = f_i({S^t_{i-1}},{S^t_i},{S^t_{i+1}})
\end{equation}
where ${S^t_{i-1}}$, ${S^t_i}$ and ${S^t_{i+1}}$ are the present states of the
left neighbor, self and right neighbor of the $i^{th}$ CA cell at $t$.

%

\begin{figure*}
\centering
\includegraphics[height=1.2in]{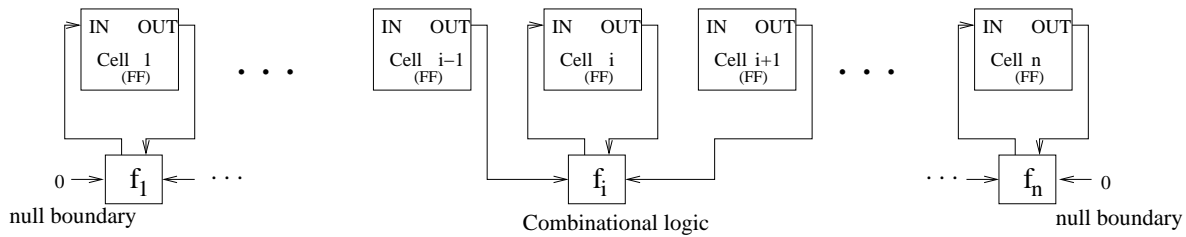}
\caption{Implementation of null boundary CA with FFs and combinational logic circuits}
\label{ca}
\end{figure*}

The collection of states ${\mathcal S}^t = (S^t_1,S^t_2,\cdots,S^t_n)$ of the cells at time $t$ is the present state of a CA.
If $S^t_0=S^t_n$ and $S^t_{n+1}=S^t_1$ (that is, left neighbor of the left most
cell is the right most cell and vice versa),
then the CA is referred to as {\em periodic boundary} CA.
On the other hand, if $S^t_0 = 0$ (null) and $S^t_{n+1} = 0$,
the CA is {\em null boundary}.
{Figure~\ref{ca}} shows the schematic diagram of a two-state 3-neighborhood null boundary CA, where each CA cell is implemented with a flip-flop ($FF$) and a combinational logic realizing
the next state function, $f_i$. This work concentrates only on null boundary CA. The examples and figures, presented in this paper point to null boundary condition.

If the next state function $f_i$ of the $i^{th}$ cell is expressed in the form of a truth table, then the decimal equivalent
of its output is conventionally referred to as the `Rule' ${\mathcal{R}}_i$ \cite{Wolfr83}.
In a two-state 3-neighborhood CA, there can be a total of $2^8$ (256) rules.
Three such rules 75, 90 and 150 are illustrated in {Table~\ref{tt}}.
{
\begin{table*}
\caption{Truth table for rule 90, 150 and 75}
\label{tt}
\[
\begin{array}{cccccccccc}
{\rm Present~ State:}    &  111 & 110 & 101 & 100 &$\underline { 011}$ &  010 &  001 &  000 & Rule \\
(RMT)& (7) & (6) & (5) & (4) & (3) & (2) & (1) & (0) & \\
{\rm ~~ (i)~ Next~ State:}    &   0 &  1  &  0  &  0  &   1  &   0  &   1  &
 1 & 75 \\
  {\rm ~(ii)~ Next~ State:}    &   0  &  1 &  0  &  1  &   1  &
   0  &   1  &   0  & 90 \\
{\rm (iii)~ Next~ State:}    &   1 &  0  &  0  &  1  &   0  &   1  &   1  &
 0 & 150 

\end{array}
\]
\end{table*}
}
The first row of the table lists the possible
2$^3$ (8) combinations of the present states of $(i-1)^{th}$, $i^{th}$
and $(i+1)^{th}$ cells at time $t$.
The last three rows indicate the next states of the
$i^{th}$  cell at $(t+1)$ for the rules, 75, 90 and 150 respectively.

From {Table~\ref{tt}}, we can also form the next state combinational logic corresponding to
a rule. That is, for
\vspace*{0.1in}
\\
{
$~~~~~~$Rule 75: $S^{t+1}_i = S^t_i.(S^t_{i-1} \oplus S^t_{i+1}) + ({\overline {S^t_{i-1}}}). ({\overline {S^t_i}})$\newline
$~~~~~~$Rule 90:  $S^{t+1}_i = S^t_{i-1} \oplus S^t_{i+1}$\newline
$~~~~~~$Rule 150: $S^{t+1}_i = S^t_{i-1} \oplus S^t_i \oplus S^t_{i+1}$.
}
\vspace*{0.1in}
\\
The next state functions $f_i$s for the rules 90 and 150 employ only $XOR$ logic.
These rules are called linear rules.
On the other hand, rule 75 is a non-linear one.
Out of total 256 rules, there are only 14 rules (15, 51, 60, 85, 90, 102, 105, 150, 153, 165, 170, 195, 204 and 240) that employ
$XOR/XNOR$ logic function and are referred to as linear/additive rules.
Other rules employ nonlinear logic functions ($AND$, $OR$, etc.).

The traditional view of CA structure was {\em uniform}, that is, the cells of a CA follow the same rule. On the other hand, in non-homogeneous or {\em hybrid} CA, the CA cells may follow different rules. For such a CA, we refer a rule vector  ${\mathcal{R}}= \langle{\mathcal{R}}_1, {\mathcal{R}}_2, \cdots, {\mathcal{R}}_i, \cdots, {\mathcal{R}}_n\rangle$, where the cell $i$ follows rule ${\mathcal{R}}_i$. Therefore, uniform CA is a special case of hybrid CA where ${\mathcal{R}}_1={\mathcal{R}}_2=\cdots={\mathcal{R}}_n$.
Whenever all the ${\mathcal{R}}_i$s ($i=1,2,\cdots,n$) of a rule vector ${\mathcal{R}}$ are
linear/additive, the CA is referred to as {\bf Linear/Additive} CA, otherwise
the CA is a {\bf Nonlinear} one. This work deals with all these form of CA (linear/additive and nonlinear; hybrid and uniform) structure under null boundary condition.

\begin{figure*}
\centering
\includegraphics[width=5.0in,height=0.9in]{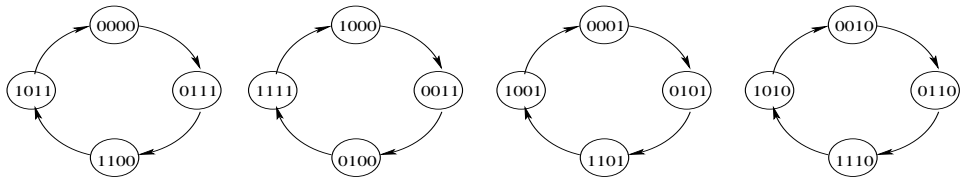}
\caption{State transitions of a reversible CA $\langle90, 15, 85, 15\rangle$}
\label{grca}
\end{figure*}

\begin{figure*}
\centering
\includegraphics[height=1.8in]{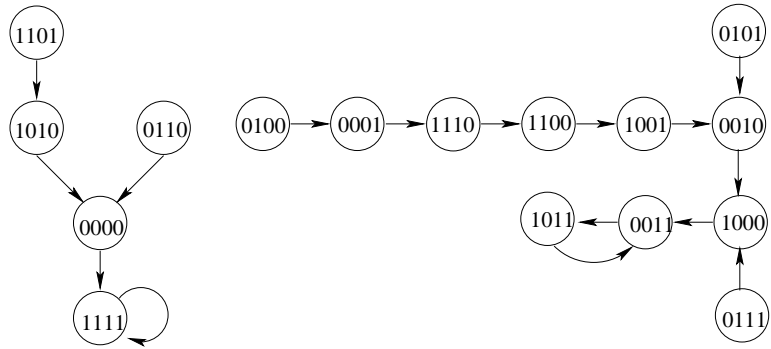}
\caption{State transitions of an irreversible CA $\langle105, 129, 171, 65\rangle$}
\label{ngrca}
\end{figure*}

The sequence of states generated (state transitions) during its evolution with time
directs the CA behavior ({Figure~\ref{grca}} and {Figure~\ref{ngrca}}).
A state transition diagram may contain {\sl cyclic} and
{\sl non-cyclic} states (a state is called {\sl cyclic} if it lies in a cycle) of a CA and
based on this, the CA can be categorized  as {\em reversible } or {\em irreversible} CA.

\begin{defn}
A CA is {\bf reversible} if it contains only cyclic states in its state transition diagram;
otherwise the CA is {\bf irreversible}.
\end{defn}

In a reversible CA, the initial CA state repeats after certain number of time steps (Figure \ref{grca}).
Therefore, all the states of a reversible CA are reachable from some other states and a state has exactly one predecessor.
On the other hand, in an irreversible CA (Figure~\ref{ngrca}),
there are some states which are not reachable ({\em non-reachable} states) from any
other state. Moreover, some states of such a CA are having more than one predecessor \cite{Moore70,Myhill63}.
For example, the states 0100 and 1101 of Figure~\ref{ngrca} are the
non-reachable states, whereas 0000 and 0010 have more than one predecessor.

The basic component of a CA structure is its cell rules. The behavior of a CA state transitions depends on
the ($i$) rules that configure the CA cells, and ($ii$) sequence of rules in ${\mathcal R}$ that forms the CA.
The next section reports characterization of the CA rules with the target to synthesize reversible CA, desired for VLSI design and test solutions.


\section{Characterization of CA rules}
\label{CArule}

This section reports the characterization of CA cell rules that helps to identify the
reversible CA structure. A tree based method is proposed for characterization of CA rules as well as
to synthesize a reversible CA in linear time.
The number of 1s in the 8-bit (8 next states) binary representation a rule plays a key role in determining the reversibility of a CA.

\begin{defn}
\label{balrule}
A rule is {\bf Balanced} if it contains equal number of 1s and 0s in
its $8-$bit binary representation; otherwise it is an {\bf Unbalanced} rule.
\end{defn}

The rules shown in Table \ref{tt} are the balanced rules.
Each of the rules has four 1s and four 0s in its 8-bit binary representation.
On the other hand, rule 171 with five 1s in its
8-bit representation (10101011) is an unbalanced rule.
In order to facilitate characterization of CA rules, we introduce further
a terminology, called {\em rule min term}.

\subsection{Rule Min Term (RMT)}
From the view point of {\em Switching Theory}, a combination of the present
states (as noted in the $1^{st}$ row of Table \ref{tt})
can be viewed as the {\sl Min Term}
of a 3-variable ($S^t_{i-1},S^t_i, S^t_{i+1}$) switching function.
Therefore, each column of the first row of
Table \ref{tt} is referred to as {\bf Rule Min Term (RMT)}.
The column 011 in the table is the RMT 3.
The next states corresponding to this RMT are 1 for Rule 75 and 90,
and 0 for Rule 150.
The characterization reported in this paper is based on analysis of the RMTs of CA rules.

\begin{defn}
\label{cmplmntRule}
A rule ${\mathcal{R}}^{'}_i$ is the {\bf complement rule} of ${\mathcal{R}}_i$ if
each RMT of ${\mathcal{R}}^{'}_i$ is the complement of the corresponding RMT of ${\mathcal{R}}_i$.
Therefore, ${\mathcal{R}}_i$ + ${\mathcal{R}}^{'}_i$ = 11111111 (255).
\end{defn}

For example, rule 90 and 165 are the complement to each other.

\noindent {\bf Relationship among RMTs:}
The RMTs of a rule dictate the next state of the CA cell, configured with that rule.
Therefore, the next state of a CA is determined by the RMTs
of all the cell rules. However, the RMTs of two consecutive cell rules
${\mathcal R}_i$ and ${\mathcal R}_{i+1}$ are related
while the CA changes its state during $t$ to $(t+1)^{th}$ instant of time.
The following discussion illustrates the relationship between the cell rules.

{
\begin{table}
\begin{center}
\caption{RMTs of the CA $\langle105, 129, 171, 65\rangle$ cell rules}
\label{rulebin}
\[
\begin{array}{cccccccccc}
RMT & 111 & 110 & 101 & 100 & 011 & 010 & 001 & 000 & $Rule$\\
	   &(7)&(6)&(5)&(4)&(3)&(2)&(1)&(0)& \\
Cell~1 & d & d & d & d & 1 & 0 & 0 & 1 & 105 \\
Cell~2& 1 & 0 & 0 & 0 & 0 & 0 & 0 & 1 & 129 \\
Cell~3 & 1 & 0 & 1 & 0 & 1 & 0 & 1 & 1 & 171 \\
Cell~4& d & 1 & d & 0 & d & 0 & d & 1 & 65 \\
\end{array}
\]
\end{center}
\end{table}
}

\begin{figure*}
\centering
\includegraphics[height=1.5in,width=4.0in]{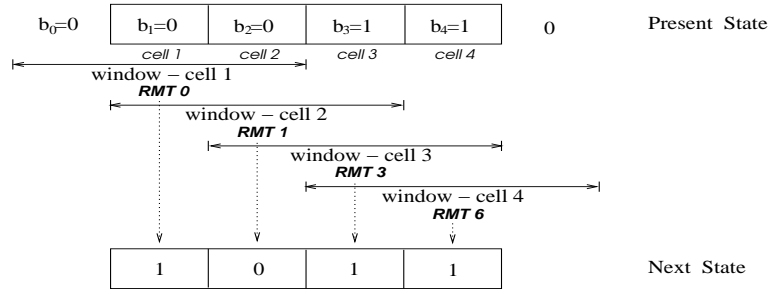}
\caption{Determining next state of a CA ($\langle105, 128, 171, 65\rangle$)}
\label{RMTwindow}
\end{figure*}

Let say, 0011 is the present state ({Figure~\ref{RMTwindow}})
of a 4-cell CA  $\langle105,129,171,65\rangle$.
The RMTs of the 4 rules are noted in {Table~\ref{rulebin}}.
As we are considering null boundary CA, the RMTs 7, 6, 5 and 4 of the left most cell rule, that is 105, are $don't~care$ (marked by $d$ in the table). Similarly, the RMTs 7, 5, 3 and 1 of the right most cell rule (65) are $don't~care$. The $don't~care$ RMTs do not appear while the CA is changing its state.

Since the CA is in 3-neighborhood, an RMT can be considered as the
3-bit window ($i-1, i, i+1$). Further, the 3-bit window
for the $(i+1)^{th}$ cell can be found from the window of $i^{th}$ cell with 1-bit right shift ({Figure~\ref{RMTwindow}}).
As the present state of the CA of {Figure~\ref{RMTwindow}} is $b_1b_2b_3b_4 =0011$,
the 3-bit window for the first cell (left most cell) is $0b_1b_2=000$.
The next state for the first cell is, therefore, guided by the RMT 0 of rule ${\mathcal R}_1=105$
 -- that is, 1 ({Table~\ref{rulebin}}).
To find the next state for second cell, the window is to be shifted right by 1-bit
position and it is $b_1b_2b_3=001$.
Hence the next state of second cell is 0 (the RMT 1 of second rule, 129).
Similarly, after 1-bit right shift, the window now becomes $b_2b_3b_4=011$.
Therefore, the next state for third cell is 1 (RMT 3 of rule 171, see {Table~\ref{rulebin}}).
Finally, the next state of the cell 4 can be computed and it is 1.
These result in CA state transition from 0011 to 1011.

It can be observed that, if the RMT window for $i^{th}$ cell is ($b_{i-1}b_ib_{i+1}$),
$b_i=0/1$, then one can predict that the RMT window for $(i+1)^{th}$ cell is either ($b_ib_{i+1}0$) or ($b_ib_{i+1}1$). In other words, if the $i^{th}$ CA cell changes
its state following the RMT $k$ (decimal equivalent of $b_{i-1}b_ib_{i+1}$) of
rule ${\mathcal R}_i$, then the $(i+1)^{th}$ cell will
generate the next state following the RMT $2k\bmod{8}$ ($b_ib_{i+1}0$)
or $(2k+1)\bmod{8}$ ($b_ib_{i+1}1$) of rule ${\mathcal R}_{i+1}$.
This relationship between the RMTs of ${\mathcal R}_i$ and ${\mathcal R}_{i+1}$
while computing the next state of a CA is shown in {Table~\ref{transition}}.
It plays an important role in characterizing the CA behavior.
We propose the concept of {\em Reachability Tree} in the following subsection
to formalize the characterization.

{
\begin{table}
\caption{Relationship between RMTs of cell $i$ and cell $(i+1)$ for next state computation}
\begin{center}
\label{transition}
\begin{tabular}{|c|c|}\hline
RMT at & RMTs at \\
$i^{th}$ rule & $(i+1)^{th}$ rule \\\hline
0 & 0, 1 \\
1 & 2, 3  \\
2 & 4, 5 \\
3 & 6, 7 \\
4 & 0, 1  \\
5 & 2, 3  \\
6 & 4, 5 \\
7 & 6, 7 \\\hline
\end{tabular}
\end{center}
\end{table}
}

\subsection{Reachability Tree}
\label{RTsec}
The Reachability tree is defined to characterize the CA states.
It is a binary tree and represents the reachable states of a CA.
Each node of the tree is constructed with RMT(s) of a rule.
The left edge of a node is considered as the 0-edge and the right edge is the 1-edge.
The number of levels of the reachability tree for an $n$-cell CA is ($n+1$).
Root node is at level 0 and the leaf nodes are at level $n$.
The nodes of level $i$ are constructed following the selected RMTs of ${\mathcal R}_{i+1}$
for the next state computation of cell $(i+1)$.
The number of leaf nodes in the tree denotes the number
of reachable states of the CA. A sequence of edges from the root to a leaf node,
representing an $n$-bit binary string,
is a reachable state, where 0-edge and 1-edge represent 0 and 1 respectively.

\begin{figure*}
\centering
\includegraphics[height=2.0in]{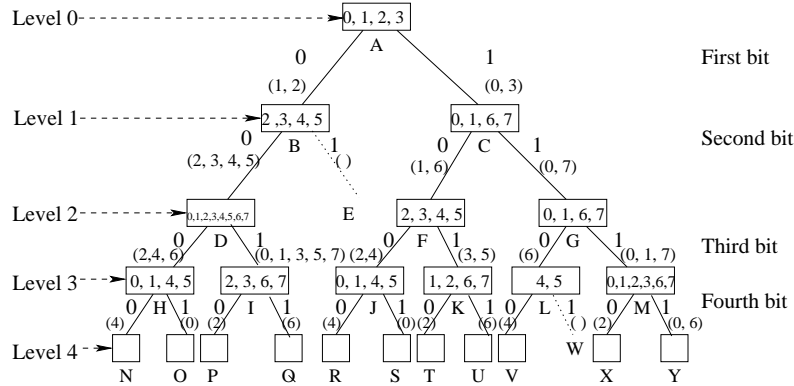}
\caption{Reachability Tree for the CA $\langle105, 129, 171, 65\rangle$}
\label{treestructure}
\end{figure*}

{Figure~\ref{treestructure}} is the reachability tree for $\langle105, 129, 171, 65\rangle$ (the RMTs of the CA rules are noted in {Table~\ref{rulebin}}).
The decimal numbers within a node at level $i$
represent the RMTs of the CA cell rule ${\mathcal R}_{i+1}$ following which
the cell $(i+1)$ may change its state.
The RMTs of a rule for which we follow 0-edge or 1-edge are noted in the bracket.
For example, the root node (level 0) is constructed with RMTs 0, 1, 2 and 3
as cell 1 can change its state following any one of the RMTs 0, 1, 2, and 3.
The rest 4, 5, 6 \& 7 are the $don't~care$ for cell 1.

For the RMTs 1 (001) and 2 (010) of 105 ({Table~\ref{rulebin}}), the next states
are 0 and it is 1 for the RMTs 0 (000) and 3 (011).
Therefore, at level 1, node after the 0-edge of level 0 contains the RMTs 2, 3, 4 \& 5
({Figure~\ref{treestructure}} and {Table~\ref{transition}}).
As the RMTs 2, 3, 4 and 5 of second cell rule (129) are 0
({Table~\ref{rulebin}}), this node does not have an 1-edge (dotted line in {Figure~\ref{treestructure}}).
It signifies that any state started with 01 (edge sequence AB, BE) is non-reachable.
On the other hand, 0010 (edge sequence AB, BD, DI, IP), 0011, etc are the reachable states
of the CA.

\begin{defn}
\label{equiD}
Two RMTs are {\bf equivalent} if both result in the same set of
RMTs effective for the next level of Reachability Tree.
\end{defn}
For example, the RMTs 0 and 4 are equivalent as both result in the same set of effective
RMTs \{000=0, 001=1\} ({Table~\ref{transition}}) for the next level of Reachability Tree.
Similarly, the RMTs 1 \& 5, 2 \& 6, and 3 \& 7 are equivalent.
\begin{defn}
\label{siblingRMT}
Two RMTs are {\bf sibling} at level $i+1$ if they are resulted in from
the same RMTs at level $i$ of the Reachability Tree.
\end{defn}
The RMTs 0 and 1 are the sibling RMTs as these two are resulted in either
from RMT 0 or from RMT 4 ({Table~\ref{transition}}).
If a node of Reachability Tree associates an RMT $k$, it also associates the sibling of $k$.

\begin{theoremm}
\label{reachBal}
The reachability tree for a reversible CA is complete.
\end{theoremm}
\begin{IEEEproof}
Since all states of a reversible CA are reachable, the number of leaf nodes in
the Reachability Tree for the $n$-cell reversible CA is $2^n$ (number of states).
Therefore, the tree is complete as it is a binary tree of ($n+1$) levels.
\end{IEEEproof}

{
\begin{table}
\begin{center}
\caption{RMTs of the CA $\langle90, 15, 85, 15\rangle$ rules}
\label{grexample}
\[
\begin{array}{cccccccccc}
RMT & 111 & 110 & 101 & 100 & 011 & 010 & 001 & 000 & $Rule$ \\
           &(7)&(6)&(5)&(4)&(3)&(2)&(1)&(0)& \\
Cell~1 & d & d & d & d & 1 & 0 & 1 & 0 & 90 \\
Cell~2& 0 & 0 & 0 & 0 & 1 & 1 & 1 & 1 & 15 \\
Cell~3 & 0 & 1 & 0 & 1 & 0 & 1 & 0 & 1 & 85 \\
Cell~4& d & 0 & d & 0 & d & 1 & d & 1 & 15 \\

\end{array}
\]
\end{center}
\end{table}
}

\begin{figure*}
\centering
\includegraphics[height=1.8in,width=3.7in]{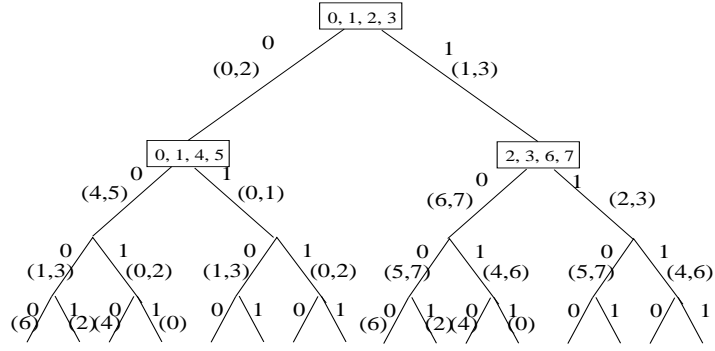}
\caption{Reachability tree for the CA $\langle90, 15, 85, 15\rangle$}
\label{reversibletree}
\end{figure*}

\begin{eqed}
\label{reversibleEx}
Let us consider a 4-cell CA $\langle90, 15, 85, 15\rangle$ noted in {Table~\ref{grexample}}. The reachability tree of the CA is shown in {Figure~\ref{reversibletree}}. The tree is a complete binary tree, and it is a reversible CA.
\end{eqed}

The above discussions point to the fact that the identification of a reversible CA (irreversible CA)
can be done by constructing the reachability tree for the CA.
If the number of non-reachable states in a Reachability Tree is zero,
then we can conclude that the CA is a reversible CA.
However, computation of the number of non-reachable states involves exponential complexity
when the CA is a reversible CA.

There is no such method to compute the number of
non-reachable states in a CA even in polynomial time.
In this work, we propose an algorithm that can identify a reversible (irreversible) CA in  $O(n)$ time.
We also report a linear time solution to synthesize an $n$-cell reversible CA.
The following theorem guides the design of such a solution.
\begin{theoremm}
\label{treeB}
The reachability tree of a 3-neighborhood null boundary CA is complete if
each edge, except the leaf edges, is resulted from exactly two RMTs of the corresponding rule.
\end{theoremm}
\begin{IEEEproof}
Let us consider an intermediate edge $l$ is resulted from a single RMT $k$ of a rule.
Now the following two cases may arise:\\
$(i)$ The edge $l$ is in between level $(n-2)$ and level $(n-1)$ (predecessor to the leaf edge):
that is, the edge $l$ connects a node of level $(n-2)$ with its one child node
at level $(n-1)$.
Therefore, the child node at level $(n-1)$ contains RMTs \{$2k\bmod{8}$, $(2k+1)\bmod{8}$\}.
Since it is a node at level $(n-1)$, the node corresponds to the CA cell rule ${\mathcal R}_n$.
As the CA is null boundary, RMT $(2k+1)\bmod{8}$ does not exist.
Hence the tree is not complete as only one edge can be generated from a single RMT.\\
$(ii)$ The edge $l$ is any intermediate edge: for this case, the very next edges
of $l$ will be resulted from RMT $2k\bmod{8}$ or from RMT $(2k+1)\bmod{8}$.
If both the RMTs are same for that particular rule, then the tree
is not complete.
Otherwise, there exist two edges and each will be resulted from a single RMT.
The process may be continued till the predecessor of the leaf node is reached.
That is, the tree may remain complete till the predecessor of the
leaf node, and there are a number of edges whose next level edge is the
leaf edge resulted from a single RMT. Hence the tree is not complete by the Case $i$.
\end{IEEEproof}

\begin{eqed}
Consider the 4-cell CA $\langle90, 15, 85, 15\rangle$ of {Example~\ref{reversibleEx}}.
The CA is a reversible CA.
Each intermediate edge of the reachability tree ({Figure~\ref{reversibletree}})
is resulted from exactly two RMTs. The RMTs are noted within the brackets.
\end{eqed}

\begin{corollaryy}
\label{4rmtNode}
All the nodes except leaves of the reachability tree for a reversible CA is
constructed with 4 RMTs.
\end{corollaryy}
\begin{IEEEproof}
Since both the 0-edge and 1-edge of a node, other than the leaves of the reachability tree for reversible CA resulted exactly from 2 RMTs ({Theorem~\ref{treeB}}), the node is, therefore, constructed with 4 RMTs.
\end{IEEEproof}

\begin{figure*}
\centering
\includegraphics[height=1.8in,width=3.7in]{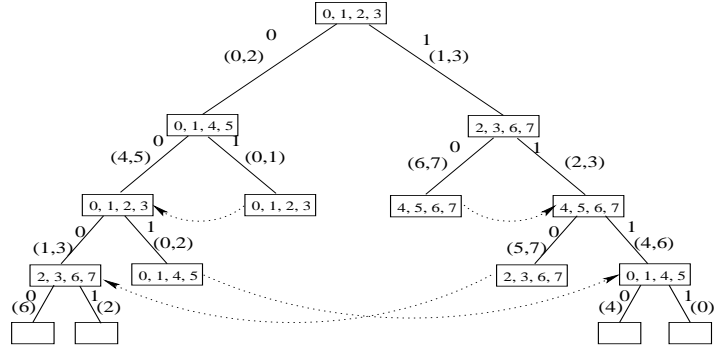}
\caption{Compressed reachability tree for the CA $\langle90, 15, 85, 15\rangle$}
\label{CRT}
\end{figure*}

There may be $2^i$ number of nodes at level $i$ of the reachability tree for an
$n$-cell CA, $i\le n$. However, all the nodes are not unique.
Two or more similar nodes at a level produce the same subtree.
The reachability tree, therefore, contains a number of similar subtrees.
For simplicity, we can show only one instance of subtree replacing other similar subtrees
of the reachability tree.
Such a reachability tree is referred to as {\em compressed reachability tree}.
{Figure~\ref{CRT}} is the compressed reachability tree of {Figure~\ref{reversibletree}}.
A dotted line points to the similar subtree.
The following theorem characterizes the nodes at each level of a reachability tree.

\begin{theoremm}
\label{uniquenode}
At each level, except root, of reachability tree for a reversible CA, there are
2 or 4 unique nodes.
\end{theoremm}
\begin{IEEEproof}
Each node of the reachability tree for a reversible CA is constructed with 4 RMTs
({Corollary~\ref{4rmtNode}})
and the sibling RMTs ({\em Definition \ref{siblingRMT}}) are associated with the same node.
Since there are 4 sets of sibling RMTs (0 \& 1, 2 \& 3, 4 \& 5, and 6 \& 7),
3 different organizations of RMTs for the nodes are possible -- \{0, 1, 2, 3\} \&
\{4, 5, 6, 7\}, \{0, 1, 4, 5\} \& \{2, 3, 6, 7\} and \{0, 1, 6, 7\} \& \{2, 3, 4, 5\}.
This implies, if a node at level $i$ is constructed with $N_1=$\{0, 1, 2, 3\}, then there exists
another node at that level constructed from $N_2=$\{4, 5, 6, 7\}.
Therefore, minimum number of unique nodes in a reachability tree of a reversible CA is 2.

It is obvious from {Theorem~\ref{treeB}} that the 2 out of 4 RMTs
({Corollary~\ref{4rmtNode}}) of a node in the reachability tree for reversible CA are $d$
($d=0/1$) and the rest 2 are $d'$.
Therefore, 2 RMTs of $N_1$ or $N_2$ are $d$, and the other 2 are $d'$.
So, another two nodes may be possible at level $i$ taking 2 RMTs that produce $d$ from $N_1$
and another 2 RMTs from $N_2$.
Hence the maximum number of possible nodes in a reachability tree for reversible CA is 4.
\end{IEEEproof}

\begin{eqed}
Consider the reversible CA of {Example~\ref{reversibleEx}}.
{Figure~\ref{CRT}} shows the unique nodes of the reachability tree for the reversible CA.
Each level except the root contains 2 unique nodes.
\end{eqed}

Based on the above discussions, we next propose a method for identification of the reversible properties
of CA followed by the synthesis scheme for an $n$-cell reversible CA in {\em Section
\ref{GrCASyn}}.

\subsection{Identification of reversible CA}
\label{GrCAAna}
This subsection proposes an algorithm (Algorithm~\ref{GrAna}) that can check whether a CA is reversible.
The algorithm scans a CA rule vector from left to right and constructs the compressed reachability tree.
It then notes an edge in the reachability tree associating other than 2 RMTs.
If there is any such edge, then the CA is irreversible ({Theorem~\ref{treeB}}).
The algorithm uses a structure $S$ with an array of sets. The number of sets
in $S$ is indicated by $nos$.
The rule vector, scanned by the algorithm, is a two dimensional array ($Rule[n][8]$),
where ($Rule[i][j]$) indicates the RMT $j$ of rule ${\mathcal R}_i$.

{
\begin{algorithm}
\caption{IdentifyReversibleCA}
\label{GrAna}
\begin{algorithmic}[1]
\REQUIRE $n$ (CA size), $Rule[n][8]$ (CA).
\ENSURE reversible or irreversible.
\STATE Find $(a)$ $S[1]=\{j\}$, where $Rule[1][j] = 0$ and $0\le j\le3$,\\
\indent \indent $(b)$ $S[2]=\{j\}$, where $Rule[1][j] = 1$ and $0\le j\le3$.\\
\indent $(c)$ If $|S[1]|\ne |S[2]|$,  report the CA as irreversible and exit.\\
\indent $(d)$ Set $nos := 2$. 
\FOR {$i=2$ to $n-1$}
\FOR {$j=1$ to $nos$}
\STATE Determine 4 RMTs for the next level node from $S[j]$ using {Table~\ref{transition}}.
\STATE Distribute these 4 RMTs into $S'[2j]$ and $S'[2j+1]$, such that $S'[2j]$ and $S'[2j+1]$ contain the RMTs that are 0 and 1
respectively for the $i^{th}$ rule.
\STATE If $|S'[2j]|\ne |S'[2j+1]|$, then report the CA as irreversible and exit.
\ENDFOR
\STATE Replace RMTs 4, 5, 6 and 7 by equivalent RMTs 0, 1, 2 and 3 respectively for each $S'[k]$.
\STATE If $|S'[k]|=1$, report the CA as irreversible and exit.
\STATE Remove duplicate sets from $S'$ and assign the sets of $S'$ to $S$.
\STATE $nos$ := number of sets in $S$.
\ENDFOR
\FOR {$j=1$ to $nos$}
\STATE Determine next 4 RMTs of $S[j]$, of which 2 are don't cares since it is the last rule.
\STATE If both the RMTs are 0 or 1 for the rule, then report the CA as irreversible and exit.
\ENDFOR
\STATE Report the CA as reversible.
\end{algorithmic}
\end{algorithm}
}


\begin{example}
\label{AnaEx}
This example illustrates the execution steps of {Algorithm~\ref{GrAna}}.
Let consider the CA $\langle90, 15,85, 15\rangle$ of {Table~\ref{grexample}}.
From {\em Step 1} of {Algorithm~\ref{GrAna}}, we get $S[1]=\{0, 2\}$ and
$S[2]=\{1, 3\}$.\\
In {\em Step 2}, when $i=2$ we obtain -- $S'[1]=\{4, 5\}$, $S'[2]=\{0, 1\}$,
$S'[3]=\{6, 7\}$ and $S'[4]=\{2, 3\}$.\\
Since each set of $S'$ contains exactly 2 RMTs, decision (reversible or irreversible) at this
stage can not be taken. Now $S'$ is modified as\\
\indent $S'[1]=\{0, 1\}$, $S'[2]=\{0, 1\}$, $S'[3]=\{2, 3\}$ and $S'[4]=\{2, 3\}$.\\
Here, each set of $S'$ contains exactly 2 RMTs. Now,
$S'$ is reduced by removing the duplicates and then assigned
to $S$. Therefore, $S[1]=\{0, 1\}$ and $S[2]=\{2, 3\}$.

\noindent When $i=3$,
$S'[1]=\{1, 3\}$, $S'[2]=\{0, 2\}$, $S'[3]=\{5, 7\}$ and $S'[4]=\{4, 6\}$.\\
Hence the modified $S'$:
$S'[1]=\{1, 3\}$, $S'[2]=\{0, 2\}$, $S'[3]=\{1, 3\}$ and $S'[4]=\{0, 2\}$.
Further, assigning reduced $S'$ to $S$, we get
$S[1]=\{1, 3\}$ and $S[2]=\{0, 2\}$.\\
Now {\em Step 3} results in
$S'[1]=\{2\}$, $S'[2]=\{6\}$, $S'[3]=\{0\}$ and $S'[4]=\{4\}$.\\
Each set of $S'$ contains a single RMT -- that is, the number of 0s and 1s in
RMTs 2, 6 and RMTs 0, 4 are the same.
So, the CA is a reversible CA ({\em Step 4}).
\end{example}

\noindent {\bf Complexity:} {\em Step 2} is the main loop in {Algorithm~\ref{GrAna}}.
It contains an inner loop ({\em Step 3}) with expected $nos$ number of iterations.
The maximum value of $nos$ is 4 as the maximum possible unique nodes at level $i$ is 4
({Theorem~\ref{uniquenode}}). Further, the loop in {\em Step 13} which also depends on $nos$.
Therefore, the execution time of the algorithm depends only on $n$.
Hence the complexity of the reversible CA identification algorithm ({Algorithm~\ref{GrAna}})
is of $O(n)$.


\subsection{Synthesis of a Reversible CA}
\label{GrCASyn}
Synthesis of reversible CA is exactly the reverse process of analysis reported in the earlier subsection.
In this subsection, we describe a reversible CA synthesis scheme in Algorithm~\ref{GrSyn}.
Input to Algorithm~\ref{GrSyn} is $n$, the size of CA to be synthesized, and the output is
an $n$-cell reversible CA. It determines the CA cell rules from analysis of the RMTs for
the desired rule ${\mathcal R}_i$, $i=1,2,\cdots,n$. The RMTs are set in such a way that
each edge of the reachability tree is resulted from two RMTs ({Theorem~\ref{treeB}}).
The algorithm also uses the two dimensional array ($Rule[n][8]$) noted in the earlier subsection.

{
\begin{algorithm}
\caption{SynthesizeReversibleCA\_1}
\label{GrSyn}
\begin{algorithmic}[1]
\REQUIRE $n$.
\ENSURE An $n$-cell reversible CA ${\mathcal R}$.
\STATE Distribute two 0s and two 1s arbitrarily in most significant 4 RMTs of $Rule[1]$.
Consider, $S[1]=\{j\}$ if $Rule[1][j] = 0$ ($0\le j\le3$) and $S[2]=\{k\}$ if
$Rule[1][k] = 1$ ($0\le k\le3$).\\
\indent Set $nos := 2$
\FOR {$i=2$ to $n-1$}
\FOR {$j=1$ to $nos$}
\STATE Determine 4 RMTs for the next level node from $S[j]$ using Table~\ref{transition}.
\STATE Distribute two 0s and two 1s arbitrarily in these 4 RMTs such that the equivalent RMTs can not be the same.
\STATE Store the RMTs that are 0 and 1 in $S'[2j-1]$ and $S'[2j]$ respectively.
\ENDFOR
\STATE Replace RMTs 4, 5, 6 and 7 by equivalent RMTs 0, 1, 2 and 3 respectively for each $S'[k]$.
\STATE Remove duplicate sets from $S'$ and assign the sets of $S'$ to $S$.
\STATE $nos$ := number of sets in $S$.
\ENDFOR
\FOR {$j=1$ to $nos$}
\STATE Determine next 4 RMTs of $S[j]$, of which 2 are don't cares since it is the last rule.
\STATE Distribute 0 and 1 randomly in the effective two RMTs for $Rule[n]$.
\ENDFOR
\STATE Report the CA as an $n$-cell reversible CA.
\end{algorithmic}
\end{algorithm}
}

\noindent {\bf Complexity:} {\em Step 2} of Algorithm~\ref{GrSyn} contains a loop that is dependent on
$n$ (number of CA cells). However, the inner loop (in {\em Step 3}) and the loop in {\em Step 12} iterate based on $nos$ (number of unique sets those derive edges).
Since the maximum value of $nos$ is constant (4), the algorithm depends only on $n$.
Hence, the complexity of Algorithm~\ref{GrSyn} is $O(n)$.

\begin{eqed}
This example illustrates the synthesis of a 4-cell reversible CA following {Algorithm~\ref{GrSyn}}.
Let us consider, two 0s and two 1s are distributed arbitrarily in least significant 4 RMTs
for the first rule ({\em Step 1}). Say, RMTs 0 and 2 are 0, that is, $S[1]=\{0, 2\}$, and the RMTs 1 and 3 are 1, that is, $S[2]=\{1, 3\}$ (Figure~\ref{R10}). Therefore, the number of nodes at level 1 is 2; one is generated from the edge that comes
from \{0, 2\} and other is from the edge that comes from \{1, 3\}.
Here, the first rule is 10. However, the RMTs for the nodes are \{0, 1, 4, 5\} and
\{2, 3, 6, 7\} {\em (Step 4)}. Two 0s and two 1s are randomly distributed in
each set so that the equivalent RMTs (RMT 0 $\&$ 4, 1 $\&$ 5, etc.) can not be the
same while these RMTs constitute a node {\em (Step 5)}.
Suppose, RMTs 0, 1, 2 and 3 are 1, and RMTs 4, 5, 6 and 7 are 0.
Hence, $S'[1]=\{4, 5\}$ $\&$ $S'[2]=\{0, 1\}$ and
$S'[3]=\{6, 7\}$ $\&$ $S'[4]=\{2, 3\}$ {(\em Step 6)}.
The $2^{nd}$ cell rule is, therefore, 15.
However, the number of sets is 4 and each set produces a node for the next level.
Since RMT 0 $\&$ 4, 1 $\&$ 5, 2 $\&$ 6, and 3 $\&$ 7 are equivalent, we replace
RMT 4 by 0, 5 by 1, 6 by 2 and 7 by 3 ({\em Step 8}).
Therefore, $S'[1]$ $\&$ $S'[4]$ and $S'[2]$ $\&$ $S'[3]$ are equivalent
and the number of unique set is 2 -- \{0, 1\} $\&$ \{2, 3\} {\em (Step 9} and {\em Step 10)}.

\begin{figure}
\centering
\includegraphics[height=0.5in]{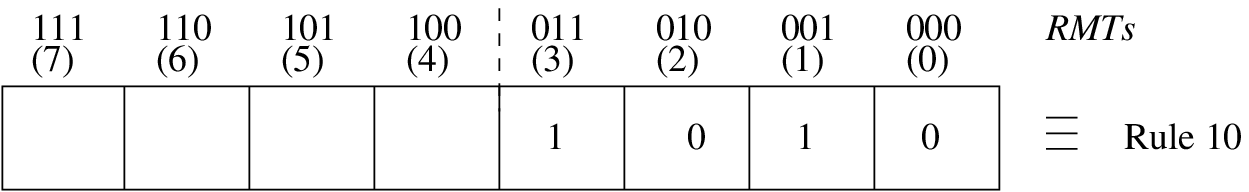}
\caption{Rule 10 as the first cell rule}
\label{R10}
\end{figure}

Similarly, the nodes for the level 2 are $S[1]=\{0, 1, 2, 3\}$ and $S[2]=\{4, 5, 6, 7\}$.
Consider, $S'[1]=\{1,3\}$ $\&$ $S'[2]=\{0, 2\}$ and $S'[3]=\{5, 7\}$ $\&$ $S'[4]=\{4, 6\}$.
That is, RMT 1, 3, 5 and 7 are 0, and RMT 0, 2, 4 and 6 are 1. Hence the $3^{rd}$ cell
rule is 85. Here, the number of unique sets is 2 (\{1, 3\} $\&$ \{0, 2\}).

The unique nodes for next level (level 3), that is, the predecessors to the leaves, are \{2, 3, 6, 7\}
and \{0, 1, 4, 5\}.
However, RMT 1, 3, 5 and 7 are don't cares, as it is the right most cell of a null
boundary CA. Therefore, the RMTs of the nodes are \{2, 6\} and \{0, 4\}.
1 and 0 are distributed randomly for the RMTs of each set.
Say, RMT 4 and 6 are 0, and RMT 0 and 2 are 1.
Then the last cell rule is 5 {(\em Step 14)}.

Therefore, the synthesized 4-cell CA is $\langle9, 15, 85, 5\rangle$.
However, the CA is equivalent to $\langle90, 15, 85, 15\rangle$ {(as that of Example~\ref{AnaEx})}.
The reachability tree and the compressed reachability tree for the CA is noted
in {Figure~\ref{reversibletree}} and {Figure~\ref{CRT}} respectively.
\end{eqed}

From {Theorem~\ref{treeB}} and {Algorithm~\ref{GrAna}}, it can be observed that each
rule of a CA plays an important role to determine the reversible/irreversible behavior of a CA.
Based on the influence of a rule in forming a reversible CA, the CA rules are further classified as {\em reversible rule} and
{\em irreversible rule}. The next section reports characterization of such CA rules.

\section{Reversible rules}
\label{grrule}
The reversible rules are the basic building blocks of a reversible CA.
Characterization of reversible rules further simplifies analysis and synthesis scheme for the reversible CA.
This section reports such characterization of CA rules and the synthesis of reversible CA \cite{Acri06}.

\begin{defnn}
A rule is a {\bf Irreversible Rule} if its presence in a rule vector makes the CA
irreversible. Otherwise, the rule is {\bf Reversible Rule}.
\end{defnn}

\begin{theoremm}
\label{unb}
An unbalanced rule is an irreversible rule.
\end{theoremm}
\begin{IEEEproof}
Let us consider ${\mathcal R}=\langle{\mathcal R}_1,{\mathcal R}_2,\cdots,
{\mathcal R}_i,\cdots,{\mathcal R}_n\rangle$ be a CA, where ${\mathcal R}_i$ is
an unbalanced rule and ${\mathcal R}{''}=\langle{\mathcal R}_1,{\mathcal R}_2,\cdots,
{\mathcal R}_i{''},\cdots,{\mathcal R}_n\rangle$ is a reversible CA.
All the rules of ${\mathcal R}$ and ${\mathcal R}{''}$ are same except the $i^{th}$ rule.
We have to prove that ${\mathcal R}$ is irreversible due to the presence of ${\mathcal R}_i$.

The reachability tree of ${\mathcal R}$ is complete up to $(i-1)^{th}$ level as
${\mathcal R}{''}$ is reversible CA with the same rules of ${\mathcal R}$ up to
$(i-1)^{th}$ cell.
Since ${\mathcal R}_i$ is unbalanced, there exists at least one node at $(i-1)^{th}$
level that has a child resulted from 1 RMT (or 3 RMTs).
This implies that the tree is not complete ({Theorem~\ref{treeB}}).
Therefore, the CA with rule vector ${\mathcal R}$ is irreversible.
Hence the proof.
\end{IEEEproof}

{\em Alternative proof:}
The above theorem can also be proved by considering the basic structure of irreversible CA state transition
diagram ({Figure~\ref{ngrca}}) that contains states with more than one predecessor.
Let us consider, $i^{th}$ rule ${\mathcal R}_i$ of a rule vector
${\mathcal R}=\langle{\mathcal R}_1,{\mathcal R}_2,\cdots,{\mathcal R}_i,\cdots,{\mathcal R}_n\rangle$
be an unbalanced rule and the
next state value of the $i^{th}$ cell corresponding to $k$ number of RMTs
of ${\mathcal R}_i$ be $d_i$, where $d_i=$0/1 and $k > 4$.
Therefore, there are $k*2^{n-3}$ number of current states
for which the next state has the form $S=\{\cdots d_i \cdots\}$.
The maximum possible number of such next states is clearly $2^{n-1}$.
Since $k*2^{n-3} > 2^{n-1}$ ($k >4$) -- that is, the
number of next states is lesser than that of current states.
It implies, there is at least a state in $S$ which contains more than one predecessor.
Therefore, the CA with unbalanced rule is an irreversible CA.

\begin{eqed}
\label{e1}
The $4-$cell CA with rule vector $\langle105, 177, 170, 75\rangle$ is a reversible CA.
Therefore, all the four rules are reversible rules.
On the other hand, a CA with rule vector $\langle105, 177, 171, 75\rangle$ is an irreversible
CA ({Figure~\ref{ngrca}}). The presence of rule 171 (in binary 10101011)
makes the CA irreversible.
That is, 171 is an irreversible rule and it is an unbalanced one.
The number of 1s in the RMTs of 171 is 5.
\end{eqed}

There are $^8C_4 = 70$ balanced CA rules in 3-neighborhood dependency.
However, all of them are not the reversible rules.
Only 62 are the reversible rules and the rest 8 are balanced irreversible rules.
The following theorem characterizes the balanced irreversible rules.

\begin{theorem}
A balanced rule with same value for the RMTs 0, 2, 3, 4, or RMTs 0, 4, 6, 7,
or RMTs 0, 1, 2, 6, or RMTs 0, 1, 3, 7 is an irreversible rule.
\end{theorem}

\begin{IEEEproof}
Let us consider, the RMTs of a balanced rule $r$ are clustered as
$g_1 = \{0, 2, 3, 4\}$ and $g_2 = \{1, 5, 6, 7\}$, where each RMT $\in$ $g_1$
is $d$ and for $g_2$ it is $d'$ ($d=0/1$). Now the following four cases may arise --

\noindent {\bf Case I:} {\underline {$r$ is the first rule of a rule vector}} --
Since RMTs 4, 5, 6 and 7 are $don't~care$s for the first rule in a null
boundary CA, the clustering of RMTs effectively becomes $g_1=$\{0, 2, 3\} and $g_2=$\{1\}.
Hence, 0-edge (1-edge) of first level of the reachability tree is resulted either from 3 or
1 (1 or 3) RMTs of $r$. Therefore, the tree is not complete ({Theorem~\ref{treeB}}).
Hence the CA with rule $r$ is irreversible -- that is, $r$ is an irreversible rule.

\noindent {\bf Case II:} {\underline {$r$ is the second rule}} --
Consider the first rule is balanced over its least significant 4 RMTs
0, 1, 2 and 3. Therefore, the possible clustering of RMTs to form the 0-edge and 1-edge
from the root can be:\\
\indent \indent\indent $f_1=[\{0, 1\} \& \{2, 3\}]$,\\
\indent \indent\indent $f_2=[\{0, 2\} \& \{1, 3\}]$, and\\
\indent \indent\indent $f_3=[\{0, 3\} \& \{1, 2\}]$.\\
That is, for $f_1$, if the RMTs 0 and 1 are 1, then the RMTs 2 and 3 are 0.
Therefore, the RMTs of level 1 nodes of the tree are \{0, 1, 2, 3\} \& \{4, 5, 6, 7\}.
Similarly, for $f_2$ it is \{0, 1, 4, 5\} \& \{2, 3, 6, 7\}, and
 \{0, 1, 6, 7\} and \{2, 3, 4, 5\} for $f_3$.
However, if the RMTs of first rule are clustered like $f_1$ or $f_3$,
the children of second level nodes are resulted from one or three RMTs. This implies, the reachability tree is not complete and the CA is irreversible.

If the RMTs of first rule are clustered as $f_2$, the RMTs of $r$ at level 1
nodes are [\{0, 4\} \& \{1, 5\}] and [\{2, 3\} \& \{6, 7\}].
Therefore, the edges of reachability tree are resulted from RMTs 0 \& 4, RMTs 1 \& 5,
RMTs 2 \& 3, and RMTs 6 \& 7. Since the RMTs 0 \& 4 (similarly 1 \& 5)
are equivalent ({\em Definition \ref{equiD}}), two nodes at level 2 are constructed with
2 RMTs. It violets {Corollary~\ref{4rmtNode}} (Section~\ref{RTsec}).
Hence the CA is irreversible.

\noindent {\bf Case III:} {\underline {$r$ is the $i^{th}$ rule}} --
Let the reachability tree of the CA is complete up to level ($i-1$).
Since RMTs 0 \& 4, 1 \& 5, 2 \& 6, and 3 \& 7 are equivalent,
without loss of generality, it can be considered that the level $i$ nodes are
generated from RMT 0, 1, 2, and 3. But any combination of these RMTs leads
to an incomplete tree (Case $II$).

\noindent {\bf Case IV:} {\underline {$r$ is the $n^{th}$ rule}} --
Since RMT 1, 3, 5 and 7 are $don't~care$s for the last rule, the clustering of RMTs
of $r$ effectively becomes $g_1 = \{0, 2, 4\}$ and $g_2 = \{6\}$.
If the reachability tree is complete up to the level ($n-1$), a number of nodes at level ($n-1$) contain two RMTs
out of 4 effective RMTs from $g_1$.
These nodes will have only a single child. This leads to an incomplete reachability tree and
the CA becomes irreversible.

This signifies that if the RMTs 0, 2, 3, 4 of a balanced rule are same, then the CA constructed with $r$ is irreversible.
Hence $r$ is an irreversible rule.
Similarly, it can also be shown that a balanced rule with same value for the RMTs \{0, 4, 6, 7\} or \{0, 1, 2, 6\} or \{0, 1, 3, 7\} is an irreversible rule. Hence the proof.
\end{IEEEproof}

\begin{corollaryy}
\label{baln}
The number of balanced irreversible CA rules in 3-neighborhood is 8.
\end{corollaryy}
\begin{IEEEproof}
As there are 4 clusterings of RMTs that lead to a balanced irreversible CA rule ({\em Theorem
\ref{e1}}) and each clustering corresponds to 2 CA rules,
the total number of such balanced irreversible rules is $4\times 2=8$.
\end{IEEEproof}

From the earlier discussion, it can be identified that the balanced irreversible rules are --
29, 46, 71, 116, 139, 184, 209 and 226. For example, the RMTs 0, 2, 3 and 4 of rule 29 (00011101) are same -- that is, 1.
The list 62 balanced reversible rules are in {Table~\ref{grruleList}}.
The 62 reversible rules can only form the reversible CA.
However, any sequence of reversible rules in a rule vector does not necessarily corresponds to a reversible CA.

\begin{table*}
\caption{List of reversible rules}
\label{grruleList}
\[
\begin{array}{|ccccccccccc|}\hline
15 & 23 & 27 & 30 & 39 & 43 & 45 & 51 &  53 &  54 &  57\\
58 & 60 & 75 & 77 & 78 & 83 & 85 &  86 & 89 &  90 &  92 \\
99 & 101& 102& 105& 106& 108& 113& 114& 120 & 135 & 141\\
142& 147& 149& 150& 153& 154& 156& 163& 165 & 166 & 169\\
170& 172& 177& 178& 180&195& 197& 198& 201& 202 & 204\\
210& 212& 216& 225& 228& 232& 240& & & & \\\hline
\end{array}
\]
\end{table*}

\begin{theoremm}
\label{seq}
Only specific sequences of reversible rules form reversible CA.
\end{theoremm}
\begin{IEEEproof}
Let us consider an $n$-cell CA designed with only reversible rules. The rules are chosen in such a way
that the CA loaded with any seed produces two types of states --
\{$\cdots d_id_{i+1} \cdots$\}
and \{$\cdots d'_id'_{i+1} \cdots$\}, where $d_i(=0/1)$ is the state of
$i^{th}$ cell and $d'_i$ is its complement. Therefore, for $2^n$ current states, the set of next states is $S=\{\cdots d_id_{i+1} \cdots, \cdots d'_id'_{i+1} \cdots \}$.
The maximum possible cardinality of $S$ is $2\times 2^{n-2}=2^{n-1}$.
Since the number of next
states is lesser than that of current states, there exists at least a
state in $S$ with more than one predecessor. Therefore, the CA is
irreversible. Hence any sequence of reversible rules can't form reversible CA.
\end{IEEEproof}

\begin{figure*}
\centering
\includegraphics[height=2.2in,width=3.8in]{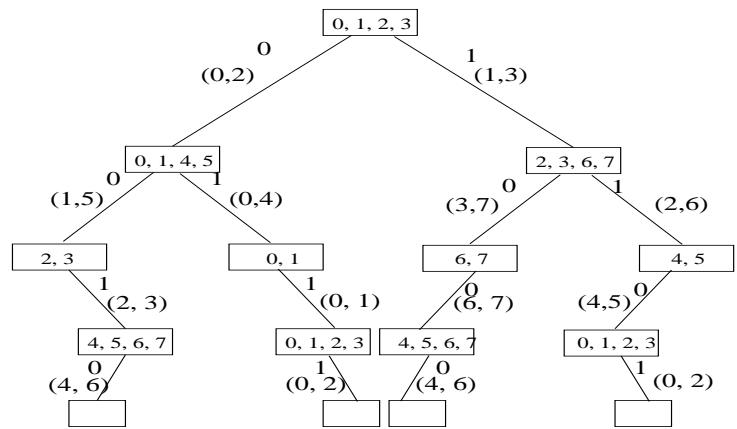}
\caption{Reachability tree for a irreversible CA $\langle90, 85, 15, 15\rangle$ designed with reversible rules}
\label{RTnon}
\end{figure*}

\begin{eqed}
The CA $\langle90, 15, 85, 15\rangle$ is a reversible CA ({Example~\ref{AnaEx}}).
However, the CA ${\mathcal R}=\langle90, 85, 15, 15\rangle$
is an irreversible CA even though each of the rules in ${\mathcal R}$ is a reversible rule {\em (Table \ref{grruleList})}.
The reachability tree for ${\mathcal R}$ is shown in {Figure~\ref{RTnon}}.
\end{eqed}

{Theorem~\ref{seq}} directs that the reversible rules are interrelated.
The sequence of reversible rules that form a reversible CA follows a specific relation.
The next section reports classification of 62 reversible rules based on the relation that must be
followed to form a rule sequence for reversible CA.

\section{Classification of reversible rules}
In the earlier section, it is reported that there are specific relations among
the reversible rules and that should be considered while synthesizing a reversible CA.
The following subsections explore such relations among the 62 reversible rules and report classification of the rules based on such relation to find the sequence of rules for a reversible CA rule vector.

\subsection{Formation of class}
\label{formclass}
Let us consider the rules ${\mathcal R}_1$, ${\mathcal R}_2$, $\cdots$, ${\mathcal R}_i$
are selected for cell 1, cell 2, $\cdots$, cell $i$ respectively to form an $n$-cell
reversible CA satisfying {Theorem~\ref{reachBal}} and {Theorem~\ref{treeB}}.
Further, consider $S$ is the set of all reversible rules ($|S|=62$).
Now, the CA cell $(i+1)$ can support a set of rules $S_j \in S$ so that any rule of $S_j$
can be selected as ${\mathcal R}_{i+1}$, satisfying the theorems {\em \ref{reachBal}} and {\em \ref{treeB}}.
We refer the class of $(i+1)^{th}$ cell as $C$ which supports
the rules of $S_j$.
The term $class$ for cell ($i+1$) as well as for the $S_j$ is used interchangeably.
Therefore, the class of $S_j$ is $C$.

\begin{lemma}
There are 6 classes of reversible CA cells in 3-neighborhood dependency.
\end{lemma}

\begin{IEEEproof}
Each node of the reachability tree of a reversible CA contains 4 RMTs ({Corollary~\ref{4rmtNode}}).
Since the sibling RMTs (Definition~\ref{siblingRMT}) are associated with the same node in the reachability tree and
there are 4 sets of sibling RMTs (0 \& 1, 2 \& 3, 4 \& 5, and 6 \& 7), 3
different organizations of RMTs for the nodes are possible.
The organizations are -- \{0, 1, 2, 3\} \& \{4, 5, 6, 7\},
\{0, 1, 4, 5\} \& \{2, 3, 6, 7\}, and \{0, 1, 6, 7\} \& \{2, 3, 4, 5\}.
Therefore, if the reachability tree contains a node with RMTs \{0, 1, 2, 3\} at $i^{th}$
level, it also contains a node with RMTs \{4, 5, 6, 7\}.

Each level of the reachability tree of a reversible CA can have either 2 or 4 unique nodes
({Theorem~\ref{uniquenode}}).
Whenever at a level, there are only 2 unique nodes, then the RMTs of the nodes may be
organized as one of the 3 possible combinations of RMTs.
For that case, the rule ${\mathcal R}_{i+1}$ is declared as of class $I$, $II$, or $III$ respectively.
On the other hand, if at the level there are only 4 unique nodes, then the RMTs of the nodes may be organized
as any two of the 3 possible combinations of RMTs.
Whenever the nodes are organized like class $I$ \& $II$, $I$ \& $III$, or $II$ \& $III$, the
class of that cell can be declared as $IV$, $V$, or $VI$ respectively.
Therefore, there are 6 classes of reversible rules.
\end{IEEEproof}

\vspace{0.1in}
\noindent {\bf Rules under each class}:
Since each node of the reachability tree for a reversible CA is constructed with 4 RMTs
({Corollary~\ref{4rmtNode}}) and both the edges (0-edge and 1-edge) of the node are resulted
from 2 RMTs ({Theorem~\ref{treeB}}), as the 2 out of 4 RMTs are 0
and others are 1.
Therefore, the RMTs of a node may be grouped as $^4C_2=6$ different ways.
However, the RMTs of the nodes of class $II$ can not be grouped as any of the 6 partitions.

For class $II$ (RMT partition is \{0, 1, 4, 5\} \& \{2, 3, 6, 7\}),
0 \& 4 (similarly 1 \& 5, 2 \& 6, and 5 \& 7)
are the equivalent RMTs ({\em Definition \ref{equiD}}) and both of these contribute same
set of RMTs for the next level.
Hence any of the equivalent RMTs are grouped together to generate a node for the next level.
That is, the number of RMTs of that node becomes 2.
This results in the CA as irreversible ({Corollary~\ref{4rmtNode}}).
Therefore, equivalent RMTs under the same node can not be grouped to produce $d$ ($d=0/1$) simultaneously.
Hence 4 groupings of RMTs out of 6 are possible in each node for class $II$.
Therefore, the number of reversible rules of class II is $4\times 4 =16$.
Since equivalent RMTs are not associated with the same node for class $I$ and $III$,
all of the 6 groupings are possible for each node.
Hence number of rules for those classes are $6\times 6=36$.
The classes and corresponding rules are given in {Table~\ref{basicclass}}.
The rules under class $IV$, $V$, and $VI$ are the common rules between $I$ \& $II$,
$I$ \& $III$, and $II$ \& $III$ respectively (as shown in the last column of Table~\ref{basicclass}, column 2 of the table notes the RMTs unique nodes).

{
\begin{table*}
\caption{Class Table}
\begin{center}
\label{basicclass}
\begin{tabular}{|c|c|c|}\hline
Class & RMTs of nodes & Rules \\\hline
I & \{0, 1, 2, 3\} & 51,  53,  54,  57,  58,  60,  83,  85,  86,\\
  & \{4, 5, 6, 7\} &      89,  90,  92,  99, 101, 102, 105, 106, 108,\\
  &                &      147, 149, 150, 153, 154, 156, 163, 165, 166,\\
  &                &      169, 170, 172, 195, 197, 198, 201, 202, 204 \\\hline
II& \{0, 1, 4, 5\} & 15,  30,  45,  60,  75,  90, 105, 120, 135,\\
   &\{2, 3, 6, 7\} & 150, 165, 180, 195, 210, 225, 240     \\\hline
III&\{0, 1, 6, 7\} & 15,  23,  27,  39,  43,  51,  77,  78,  85,\\
   &\{2, 3, 4, 5\} & 86,  89,  90, 101, 102, 105, 106, 113, 114,\\
   &               & 141, 142, 149, 150, 153, 154, 165, 166, 169,\\
   &               & 170, 177, 178, 204, 212, 216, 228, 232, 240 \\\hline
IV &\{0, 1, 2, 3\} & 60,  90, 105, 150, 165, 195            \\
   &\{4, 5, 6, 7\} &		\\
   &\{0, 1, 4, 5\} &		\\
   &\{2, 3, 6, 7\} &		\\\hline
V  &\{0, 1, 2, 3\} & 51,  85,  86,  89,  90, 101, 102, 105, 106, 149, \\
   &\{4, 5, 6, 7\} & 150, 153, 154, 165, 166, 169, 170, 204 \\
   &\{0, 1, 6, 7\} &		\\
   &\{2, 3, 4, 5\} &		\\\hline
VI &\{0, 1, 4, 5\} & 15,  90, 105, 150, 165, 240              \\
   &\{2, 3, 6, 7\} &		\\
   &\{0, 1, 6, 7\} &		\\
   &\{2, 3, 4, 5\} &		\\\hline

\end{tabular}
\end{center}
\end{table*}
}

%
%

\begin{figure*}
\begin{center}
\centering
\includegraphics[width=4.2in,height=1.9in]{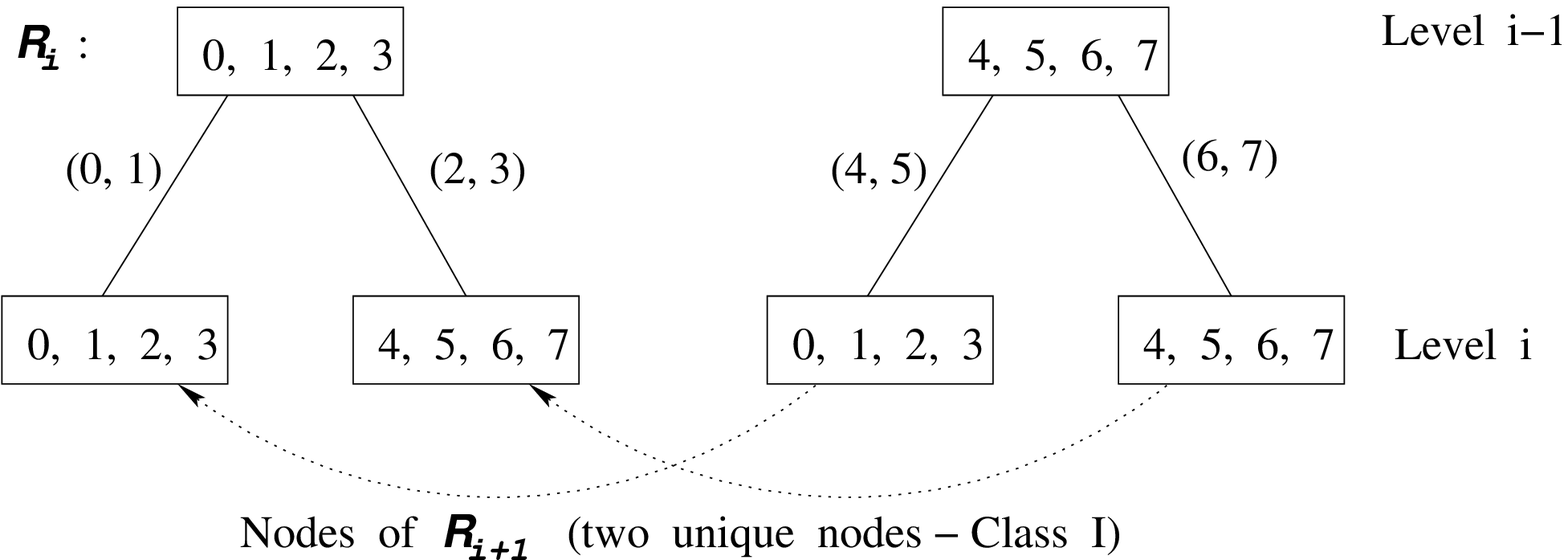}
\vspace{0.1in}
\mbox{ ($a$) Next rule class is I}
\vspace{0.4in}
\centering
\includegraphics[width=4.2in,height=1.9in]{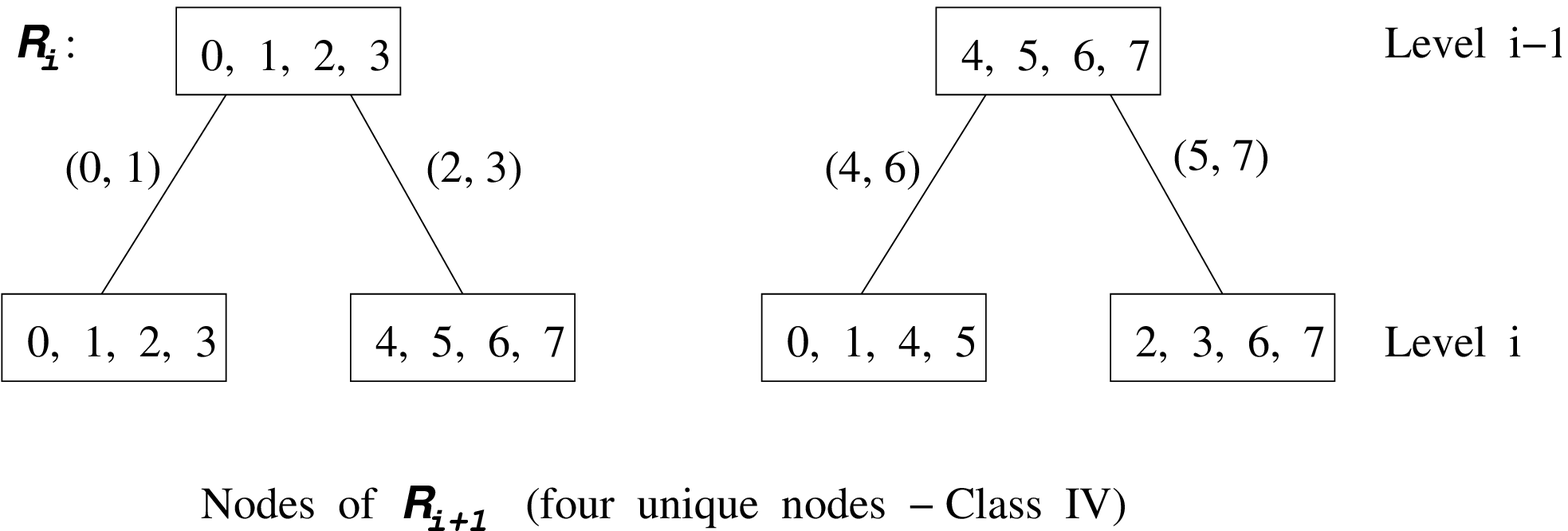}
\mbox{ ($b$) Next rule class is IV}
\caption{Determination of class relationship}
\label{DetermineClass}
\end{center}
\end{figure*}

\subsection{Class relationship between ${\mathcal R}_i$ and ${\mathcal R}_{i+1}$}
\label{ClassRelation}
This section determines the relationship between the classes of ${\mathcal R}_i$ and
${\mathcal R}_{i+1}$. From the known
${\mathcal R}_i$ and its class, we can find the class of ${\mathcal R}_{i+1}$ \cite{Acri06}.

Let us consider the class of ${\mathcal R}_i$ be $I$ ({Figure~\ref{DetermineClass}}).
Therefore, two unique nodes having RMTs \{0, 1, 2, 3\} and \{4, 5, 6, 7\} are available at the $(i-1)^{th}$ level of
the reachability tree. Now consider the RMTs of ${\mathcal R}_i$ are clustered as
\{0, 1, 4, 5\} and \{2, 3, 6, 7\}, where the RMTs of a set are the same, that is, either 0 or 1.
In {Figure~\ref{DetermineClass}(a)}, the RMTs \{0, 1, 4, 5\} are considered as 0, and
the RMTs \{2, 3, 6, 7\} as 1.
Therefore, the RMTs are grouped as (0, 1), (2, 3), (4, 5) and (6, 7).
Each edge of the nodes is resulted from any one of these groups.
Hence two edges connecting the node ($N_1$) having RMTs \{0, 1, 2, 3\} with its children are resulted
from (0, 1) and (2, 3).
Therefore, the two children $N_1C_1$ and $N_1C_2$ (for next level) of $N_1$ are having RMTs
\{0, 1, 2, 3\} and \{4, 5, 6, 7\} ({Table~\ref{transition}}) ({Figure~\ref{DetermineClass}(a)}).
Similarly, the children $N_2C_1$ and $N_2C_2$ of the node $N_2$ with RMTs
\{4, 5, 6, 7\} are constructed with RMTs \{0, 1, 2, 3\} and \{4, 5, 6, 7\} --
that is, the nodes $N_2C_1$ and $N_2C_2$ are same with the other two children ($N_1C_1$ and $N_1C_2$ of $N_1$).
Therefore, the next level of the reachability tree contains two unique nodes
having RMTs \{0, 1, 2, 3\} and \{4, 5, 6, 7\} ({Figure~\ref{DetermineClass}(a)}).
Hence the class of ${\mathcal R}_{i+1}$ is $I$.

Further, if the RMTs of ${\mathcal R}_i$ are grouped as (0, 1), (2, 3), (4, 6), and (5, 7)
({Figure~\ref{DetermineClass}(b)}),
the nodes of $N_1C_1$ and $N_1C_2$ level $i$, generated from the node $N_1$ of level $(i-1)$ with RMTs
\{0, 1, 2, 3\}, are having RMTs \{0, 1, 2, 3\} and \{4, 5, 6, 7\}.
The other two nodes ($N_2C_1$ and $N_2C_2$) at level $i$, generated from the node $N_2$ with RMTs \{4, 5, 6, 7\}, are
having RMTs \{0, 1, 4, 5\} and \{2, 3, 6, 7\}.
In this case, the next level of reachability tree contains four unique nodes having RMTs
\{0, 1, 2, 3\}, \{4, 5, 6, 7\}, \{0, 1, 4, 5\}, and \{2, 3, 6, 7\} ({Figure~\ref{DetermineClass}(b)}).
Therefore, the organizations
of RMTs support the property of both the Class $I$ and Class $II$ -- that is, Class IV (Table~\ref{basicclass}, column 2). So, the class of ${\mathcal R}_{i+1}$ is $IV$.

{
\begin{table*}
\caption{Formation of class relationship between ${\mathcal R}_i$ and ${\mathcal R}_{i+1}$}
\begin{center}
\label{classrelation}
\begin{tabular}{|c|c|c|c|c|}\hline
(1)	& (2)	& (3)	& (4)	& (5)	\\
Class & RMTs of & Groupings of & RMTs of & Class \\
 of &  unique nodes & RMTs & unique nodes &  of\\
${\mathcal R}_i$ &at level $(i-1)$ & at level $(i-1)$ & at level $i$ & ${\mathcal R}_{i+1}$ \\\hline
I & \{0, 1, 2, 3\} & (0, 1), (2, 3) & \{0, 1, 2, 3\} & I\\
  & \{4, 5, 6, 7\} & (4, 5), (6, 7) & \{4, 5, 6, 7\} & \\\cline{3-5}
  &     	   & (0, 2), (1, 3) & \{0, 1, 4, 5\} & II \\
  &     	   & (4, 6), (5, 7) & \{2, 3, 6, 7\} &  \\\cline{3-5}
  &     	   & (0, 3), (1, 2) & \{0, 1, 6, 7\} & III \\
  &     	   & (4, 7), (5, 6) & \{2, 3, 4, 5\} &  \\\cline{3-5}
  &     	   & \{(0, 1), (2, 3) & \{0, 1, 2, 3\} & IV \\
  &     	   & (4, 6), (5, 7)\} & \{4, 5, 6, 7\} &  \\
  &     	   & or \{(0, 2), (1, 3) & \{0, 1, 4, 5\} &  \\
  &     	   & (4, 5), (6, 7)\} & \{2, 3, 6, 7\} &  \\\hline
II& \{0, 1, 4, 5\} & (0, 1), (4, 5) & \{0, 1, 2, 3\} & I\\
   &\{2, 3, 6, 7\} & (2, 3), (6, 7) & \{4, 5, 6, 7\} & \\\hline
IV & \{0, 1, 2, 3\} & (0, 1), (2, 3) & \{0, 1, 2, 3\} & I\\
   & \{4, 5, 6, 7\} & (4, 5), (6, 7) & \{4, 5, 6, 7\} & \\\cline{3-5}
   & \{0, 1, 4, 5\} & \{(0, 1), (2, 3) & \{0, 1, 2, 3\} & IV \\
   & \{2, 3, 6, 7\} & (4, 6), (5, 7)\} & \{4, 5, 6, 7\} & \\
   &		    & or \{(0, 2), (1, 3) & \{0, 1, 4, 5\} &\\
   &		    & (4, 5), (6, 7)\}    & \{2, 3, 6, 7\} &\\\hline
\end{tabular}
\end{center}
\end{table*}
}

{Table~\ref{classrelation}} displays the relationship (partly) between reversible rules.
Only 3 classes, $I$, $II$, and $IV$ are selected to illustrate the relationship.
First column shows the class of ${\mathcal R}_i$. Column 2 notes the RMTs of unique nodes at level $(i-1)$.
Whereas, Column 3 shows the grouping of RMTs for ${\mathcal R}_i$.
The RMTs of unique nodes at level $i$ are shown in Column 4. Based on the unique nodes
at level $i$, the class of ${\mathcal R}_{i+1}$ is decided and is reported in Column 5.

The details of relationship among the classes are reported in {Table~\ref{nextclass}}.
The first and second columns of the table represent the class of $i^{th}$
cell and the rule ${\mathcal R}_i$ respectively, whereas the class of the $(i+1)^{th}$ cell
corresponding to this pair (the class of $i^{th}$ cell and ${\mathcal R}_i$) is noted in
the last column.
It can be observed that a rule can be the member of more than one class. For example,
rule 90, 105, 150 and 165 are the members of all the 6 classes.
Such rules are referred to as the complete rules.

{
\begin{table*}
\caption{Class relationship of ${\mathcal R}_i$ and ${\mathcal R}_{i+1}$}
\begin{center}
\label{nextclass}
\begin{tabular}{|c|c|c|}\hline
Class of & ${\mathcal R}_i$ & Class of\\
${\mathcal R}_i$ & & ${\mathcal R}_{i+1}$ \\\hline
I & 51,  60,  195, 204 & I\\\cline{2-3}
  &                       85,  90,  165, 170 & II \\\cline{2-3}
  &     102, 105, 150, 153 & III \\\cline{2-3}
  &     53, 58, 83, 92, 163, 172, 197, 202 & IV \\\cline{2-3}
  &     54, 57, 99, 108, 147, 156, 198,201 & V \\\cline{2-3}
  &     86, 89, 101, 106, 149, 154, 166, 169 & VI \\\hline
II& 15,  30,  45,  60,  75,  90, 105, 120, 135, & I \\
   &                      150, 165, 180, 195, 210, 225, 240 & \\\hline
III& 15,  51, 204, 240 & I \\\cline{2-3}
   &                       85, 105, 150, 170 & II\\\cline{2-3}
   &    90, 102, 153, 165 & III \\\cline{2-3}
   &    23, 43, 77, 113, 142, 178, 212, 232 & IV\\\cline{2-3}
   &    27, 39, 78, 114,  141, 177, 216, 228 & V\\\cline{2-3}
   &    86, 89, 101, 106, 149, 154, 166, 169 & VI \\\hline
IV & 60, 195 & I \\\cline{2-3}
   & 90, 165 & IV \\\cline{2-3}
   & 105, 150 & V \\\hline
V  & 51, 204 & I\\\cline{2-3}
   & 85, 170 & II\\\cline{2-3}
   & 102, 153 & III\\\cline{2-3}
   & 86, 89, 90, 101, 105, 106, 149, 150, & VI\\
   &154, 165,166, 169 & \\\hline
VI & 15, 240 & I\\\cline{2-3}
   & 105, 150 & IV \\\cline{2-3}
   & 90, 165 & V \\\hline
\end{tabular}
\end{center}
\end{table*}
}

\begin{defn}
\label{completeRule}
A rule is {\bf complete} if it is the member of all the six classes.
\end{defn}

\vspace{0.1in}
\noindent {\bf First and Last rule}: The class identification of rules is applicable for both
the null boundary and periodic boundary CA.
In this work, we have concentrated only on 1-dimensional 3-neighborhood null boundary
CA.

For null boundary CA, the RMTs 4, 5, 6 and 7 are the {\sl don't cares} for
${\mathcal R}_1$ (left most cell rule) as the present state of left neighbor of cell 1 (left most cell of a CA) is always 0.
So, there are only 4 {\em effective} RMTs (0, 1, 2, 3) for
${\mathcal R}_1$.
Similarly, the RMTs 1, 3, 5 and 7 are the don't care RMTs for
${\mathcal R}_n$ (right most cell rule).
The {\em effective} RMTs for ${\mathcal R}_n$ are, therefore, 0, 2, 4 and 6.
That is, rule 105 and 9 are equivalent if selected for the ${\mathcal R}_1$.
Similarly, the rules 75 and 65 are effectively the same while chosen for the $n^{th}$ CA cell.
Therefore, there are $2^{2^2}=16$ effective rules for the ${\mathcal R}_1$ as well as for the ${\mathcal R}_n$.

\begin{corollaryy}
\label{Cfl}
If ${\mathcal R}=\langle{\mathcal R}_1,{\mathcal R}_2,\cdots,{\mathcal R}_n\rangle$ is
a reversible CA, then ${\mathcal R}_1$ and
${\mathcal R}_n$ are balanced over their effective 4 RMTs.
\end{corollaryy}

\begin{IEEEproof}
Let us consider, the first rule is unbalanced over its 4 effective RMTs.
That is, the next state of 3 RMTs out of 4 effective RMTs of ${\mathcal R}_1$ be $d$ ($d=0/1$).
Therefore, there are $3*2^{n-2}$ number of current states for which the next state has the form
$S = \{d\cdots\}$. The maximum possible number of such next states is clearly $2^{n-1}$.
Since the number of next states is lesser than that of current states, there is at least a state
in $S$ which contains more than one predecessor. Hence the CA is irreversible.
This is because of that the ${\mathcal R}_1$ is unbalanced over its 4 effective RMTs.
Therefore, to form a reversible CA, ${\mathcal R}_1$ must be balanced over its 4 effective RMTs.
With similar logic, it can be proved that ${\mathcal R}_n$ has to be balanced over its 4 effective RMTs.
\end{IEEEproof}

Corollary~\ref{Cfl} signifies that the unbalanced rule 3 is a reversible rule when it is selected as the ${\mathcal R}_1$.
The rule 3 ${\mathcal R}_1$) as is balanced over its 4 effective RMTs.
There are $^4C_2=6$ rules (out of total 16 effective rules for the ${\mathcal R}_1$)
that are balanced over their 4 effective RMTs.
{Table~\ref{first}} identifies such 6 rules
and the corresponding class of rule ${\mathcal R}_2$ for the CA cell 2.
The similar consideration is also true for the ${\mathcal R}_n$.
{Table~\ref{last}} lists all such 6 reversible rules for the ${\mathcal R}_n$.

{
\begin{table}
\caption{First Rule Table}
\begin{center}
\label{first}
\begin{tabular}{|c|c|c|c|}\hline
Rules for & group & RMTs of nodes & Class of \\
${\mathcal R}_1$ & of RMTs & for level 2 & ${\mathcal R}_2$ \\\hline
 3, 12 & (0, 1) &\{0, 1, 2, 3\} &I\\
	&(2, 3)	&\{4, 5, 6, 7\} & \\\hline
 5, 10 & (0, 2) &\{0, 1, 4, 5\} &II\\
	&(1, 3)	&\{2, 3, 6, 7\} & \\\hline
 6, 9 & (0, 3) &\{0, 1, 6, 7\} &III\\
	&(1, 2)	&\{2, 3, 4, 5\} & \\\hline
\end{tabular}
\end{center}
\end{table}
\begin{table}
\caption{Last Rule Table}
\begin{center}
\label{last}
\begin{tabular}{|c|c|}\hline
Rule class & Rule set\\
for ${\mathcal R}_n$ & for ${\mathcal R}_n$ \\\hline
 I & 17, 20, 65, 68 \\
  II & 5, 20, 65, 80 \\
 III & 5, 17, 68, 80 \\
 IV & 20, 65 \\
 V & 17, 68 \\
 VI & 5, 80 \\\hline
\end{tabular}
\end{center}
\end{table}
}
 

\subsection{Reversible CA synthesis}
A reversible CA synthesis scheme is proposed in {Algorithm~\ref{GrSyn}} ({Section~\ref{GrCASyn}}).
In this subsection, we develop a relatively simpler method to synthesize a reversible CA exploiting the class relationships of reversible CA rules presented in the tables \ref{nextclass}, \ref{first} and \ref{last}.
For example, let us consider the synthesis of a 4-cell reversible CA and say,
rule 9 is selected randomly as ${\mathcal R}_1$ from {Table~\ref{first}}.
Therefore, the class (obtained from {Table~\ref{first}}) of
$2^{nd}$ cell rule is III. From Class III of {Table~\ref{nextclass}}, say rule
177 is selected randomly as the ${\mathcal R}_2$.
Therefore, the class of ${\mathcal R}_3$ is found to be V, since
the class of $2^{nd}$ cell rule is III and ${\mathcal R}_2=177$ ({Table~\ref{nextclass}}).
We select rule 170 as ${\mathcal R}_3$ from Class V of {Table~\ref{nextclass}}. The class of last ($4^{th}$) cell is, therefore, II.
Rule 65 is selected randomly for ${\mathcal R}_4$
from {Table~\ref{last}}. Therefore, the 4-cell reversible CA is
${\mathcal R} = \langle9, 177, 170, 65\rangle$.
The formal algorithm, of $O(n)$ complexity, to synthesize a reversible CA is presented below.

\begin{algorithm}
\caption{SynthesizeReversibleCA\_2}
\label{Synthesis}
\begin{algorithmic}[1]
\REQUIRE $n$ (length of $CA$), tables \ref{nextclass},
\ref{first} and \ref{last}.
\ENSURE A reversible CA -- that is, the rule vector ${\mathcal {R}} = <{\mathcal {R}}_1, {\mathcal {R}}_2, \cdots, {\mathcal {R}}_n>$.
\STATE Pick up the first rule ${\mathcal R}_1$ randomly from
{Table~\ref{first}} 
\STATE  $cl  \Leftarrow $ Class of ${\mathcal R}_2$ ($cl \in$ \{I, II, III\})
\FOR {$i:=2$ to $n-2$}
\STATE From class $cl$ of Table \ref{nextclass}, select ${\mathcal R}_{i}$, satisfying $C_2$, randomly.
\STATE Find $cl$ for the $(i+1)^{th}$ cell rule from Table \ref{nextclass} ($cl \in$ \{I, II, III, IV, V, VI\}) based on the ${\mathcal R}_i$ and its class.
\ENDFOR
\STATE From class $cl$ of {Table~\ref{last}}, pick up a rule as ${\mathcal R}_n$.
\STATE Form the rule vector ${\mathcal {R}} = <{\mathcal {R}}_1, {\mathcal {R}}_2, \cdots, {\mathcal {R}}_n>$.
\end{algorithmic}
\end{algorithm}

\noindent {\bf Complexity:} Algorithm~\ref{Synthesis} utilizes a single $for$ loop in {\em Step 3}, which depends on the value of $n$ (CA size). Obviously, the complexity of the algorithm is $O(n)$.


\section{Conclusion}
This paper reports the detail characterization of 1-dimensional 3-neighborhood non-homogeneous/hybrid
CA under null boundary condition. The concept of reachability tree is introduced to characterize the CA.
An $O(n)$ time solution scheme is proposed to decide on the reversibility of a CA.
A linear time solution is also proposed for the synthesis of reversible CA through
classification of all the 256 CA rules into 6 classes.

\bibliographystyle{plain}
\bibliography{References}

\end{document}